\documentclass[12pt]{article}
\usepackage{jcapmod}
\usepackage{rotating}
\usepackage{amsmath}
\usepackage{mathtools}
\usepackage{amsthm}
\usepackage{amsfonts}
\usepackage{graphicx}
\usepackage{psfrag}
\usepackage{hyperref}
\usepackage{amssymb}
\usepackage{lscape}
\usepackage{slashed}
\usepackage{lscape}
\usepackage{color}
\usepackage{url}
\usepackage{caption}
\usepackage{bm}
\usepackage{subfig}
\usepackage{epsfig}
\usepackage{tikz}
\usepackage{tikz-cd}

\DeclareMathOperator{\csch}{csch}

\definecolor{teal}{rgb}{0.1,0.5,0.1}

\hypersetup{
    colorlinks = true,
    citecolor = {teal},
    linkcolor = {teal},
    urlcolor = {teal},
}

\def\ga{\mathrel{\raise.3ex\hbox{$>$\kern-.75em\lower1ex\hbox{$\sim$}}}}
\def\la{\mathrel{\raise.3ex\hbox{$<$\kern-.75em\lower1ex\hbox{$\sim$}}}}

\def\beq{\begin{equation}}
\def\eeq{\end{equation}}
\def\bea{\begin{eqnarray}}
\def\eea{\end{eqnarray}}

\newcommand{\lslashslash}{%

  \raisebox{0.8ex}{%
    \scalebox{.7}{%
      \rotatebox[origin=c]{18}{$-$}%
    }%
  }%
}
\newcommand{\lslash}{%
  {%
   \vphantom{d}%
   \ooalign{\kern-.1em\smash{\lslashslash}\hidewidth\cr${\rm l}$\cr}%
   \kern.05em
  }%
}

\begin{document}

\begin{flushright}
UMN--TH--4019/21, FTPI--MINN--21/12, IFT-UAM/CSIC-21-83 
\end{flushright}

\vspace{0.5cm}
\begin{center}
{\bf {\large On the Realization of WIMPflation}}
\end{center}

\vspace{0.05in}

\begin{center}{
{\bf Marcos A.~G.~Garcia}$^{a,b}$,
{\bf Yann Mambrini}$^{c}$},
{\bf Keith~A.~Olive}$^{d}$, and
{\bf Sarunas Verner}$^{d}$
\end{center}

\begin{center}
{\em $^{a}$ Instituto de F\'isica Te\'orica (IFT) UAM-CSIC, Campus de Cantoblanco, 28049, Madrid, Spain}\\[0.2cm] 
{\em $^{b}$ Departamento de F\'isica Te\'orica, Universidad Aut\'onoma de Madrid (UAM), Campus de Cantoblanco, 28049 Madrid, Spain}\\[0.2cm] 
  {\em $^c$ Universit\'e Paris-Saclay, CNRS/IN2P3, IJCLab, 91405 Orsay, France}\\[0.2cm] 
 {\em $^d$William I. Fine Theoretical Physics Institute, School of
 Physics and Astronomy, University of Minnesota, Minneapolis, MN 55455,
 USA}\\[0.2cm]

\end{center}

\bigskip

\centerline{\bf ABSTRACT}

\noindent  
We consider models for inflation with a stable inflaton. Reheating is achieved through scattering processes such as $\phi \phi \to h h$,
where $h$ is the  Standard Model Higgs boson. We consider the reheating process in detail and show that for a relatively large coupling (needed for the late annihilations of the inflaton during freeze-out), reheating is almost instantaneous leading to a relatively high reheating temperature. The process $\phi \phi \leftrightarrow h h$ brings the inflaton back into equilibrium, leading to a well studied scalar singlet dark matter candidate and Higgs portal model. We argue that such models can be derived from no-scale supergravity. 

\vspace{0.2in}

\begin{flushleft}
July 2021
\end{flushleft}
\medskip
\noindent


\newpage

\section{Introduction}

There are many aspects of cosmology which rely on physics beyond the Standard Model. These include inflation and a dark matter candidate as well as other problems such as baryogenesis and dark energy. In most models of inflation \cite{reviews}, inflation is driven by one (or more) scalar fields. In single-field inflation models,
this field, barring a direct connection to a specific grand unified theory (GUT) as in early models of new inflation \cite{new}, may be a generic scalar field dubbed the inflaton \cite{nos}. Generally,
the inflaton is not stable and its decays lead to the reheating and thermalization of the universe after the period of exponential expansion \cite{dg,nos}. In this case, the only connection between inflation and dark matter, is the role of the inflaton in producing dark matter during reheating, either through the process of thermalization  \cite{Davidson:2000er,Harigaya:2013vwa,Mukaida:2015ria,Garcia:2018wtq} when reheating is not assumed to be instantaneous \cite{Giudice:2000ex,Garcia:2017tuj,Chen:2017kvz,Garcia:2020eof,Bernal:2020gzm,Co:2020xaf,Garcia:2020wiy} or through the direct decay \cite{grav2,grav3,egnop,Kaneta:2019zgw} 
or scattering \cite{Mambrini:2021zpp} of the inflaton to dark matter. These mechanisms are particularly important for dark matter candidates which never achieve thermal equilibrium with the  radiation background, such as the gravitino \cite{ehnos,kl,nos}, or more generic feebly-interacting massive particles (FIMPs) \cite{fimp,Bernal,Bernal:2017kxu,Chen:2017kvz,Bernal:2019mhf,Bernal:2020gzm}. 
Furthermore, the evolution of the temperature 
during reheating also plays a role for dark matter produced directly from 
inflaton decays, either at the tree level \cite{egnop,Garcia:2017tuj,grav2,Garcia:2020eof}, or at the loop level \cite{Kaneta:2019zgw}. 

If, however, the inflaton is protected by a symmetry and hence stable, it may be a candidate for dark matter \cite{Liddle:2006qz,Lerner:2009xg,delaMacorra:2012sb,Mukaida:2014kpa,Bastero-Gil:2015lga,hooper,Rosa:2018iff,Almeida:2018oid,kamion,oleg}. Of course, if the inflaton is stable, reheating requires an alternative mechanism,
which may be provided perhaps by non-perturbative effects or a coupling to curvature, but generally these are difficult to achieve. Gravitational particle production caused by a changing space-time metric is usually inefficient and may lead to a relatively low reheating temperature that is incompatible with Big Bang Nucleosynthesis~\cite{parker, zeldovich, ford}, and reheating through non-perturbative effects is impossible to achieve unless additional interaction terms are introduced that would make the inflaton unstable and not suitable dark matter candidate~\cite{preheating,preheating2}.

For an inflaton potential dominated by a quadratic term,
with a mass scale which is determined by the normalization of the Cosmic Microwave Background (CMB) anisotropy spectrum, $m_\phi \simeq 3 \times 10^{13}$ GeV, inflaton scattering (e.g., $\phi \phi \to h h$, where $h$ is the Standard Model Higgs boson),
does not lead to reheating in the absence of reheating through inflaton decays.\footnote{We define the moment of reheating when the energy density in radiation, $\rho_R (a_{\rm RH})$,  is equal to the energy density stored in inflaton oscillations $\rho_\phi (a_{\rm RH})$, where $a_{\rm RH}$ is the cosmological scale factor when the two energy densities are equal. Furthermore, successful reheating requires that $\rho_R (a) > \rho_\phi (a)$ for $a > a_{\rm RH}$.} In this case, 
scatterings lead to $\rho_R \propto a^{-4}$ compared with $\rho_\phi \propto a^{-3}$, and reheating does not occur \cite{Garcia:2020wiy} 
(this will be discussed in more detail below). 
In contrast, radiation produced by inflaton decays
falls as $\rho_R \propto a^{-3/2}$ and definitely leads to reheating. For sufficiently large couplings, preheating \cite{preheating, preheating2, stb, kt,Greene:1997fu} can
produce radiation early in the oscillatory phase, and this may be important for massive particle production, but this alone does not lead to reheating and a radiation-dominated universe \cite{preheating2,gkmov}. 

The evolution of the temperature during the reheating process, however, depends on the shape of the inflaton potential about the minimum \cite{Bernal:2019mhf,Garcia:2020eof,Garcia:2020wiy}. 
For example, for a potential which behaves as
\beq
V \; =  \; \lambda \left( \frac{\phi^k}{M_P^{k-4}}\right) \, ,
\label{Vk}
\eeq
the evolution of the temperature of the radiation produced by scatterings during oscillations about the minimum is given by \cite{Garcia:2020eof,Garcia:2020wiy}
\footnote{Note that this dependence is altered when the effective mass of the final state bosons is much greater than the effective mass of the inflaton as we discuss in more detail in Section \ref{sec:RH}.}
\beq
T \; \propto \; a^{-\frac{9}{2k+4}} \, ,
\eeq
so that a potential dominated by a quartic term leads to 
$\rho_R \propto a^{-3}$ whereas $\rho_\phi \propto a^{-4}$ in this case, and reheating becomes possible. As a result, 
a potential of the form
\beq
V \; = \; \frac12 m_\phi^2 \phi^2 + \lambda \left( \frac{\phi^k}{M_P^{k-4}}\right) + \sigma \phi^2 b^2 \, ,
\label{Vmk}
\eeq
where $m_\phi$ is a weak scale mass and the dynamics of inflation
is determined by $\lambda$, can lead to inflation, reheating through the coupling $\sigma$, and dark matter with a weak scale mass, $m_\phi$.  We emphasize that reheating in this context is only possible when $m_\phi^2 \ll \lambda \phi^{k-2}$ so that the evolution of the radiation bath is determined by $\lambda$ rather than $m_\phi$.

While many models of inflation including the Starobinsky model~\cite{Starobinsky:1980te,Mukhanov:1981xt,Starobinsky:1983zz}
lead to quadratic inflaton oscillations, this is not a necessary feature of inflationary models. For example,
some $\alpha$-attractor models, in particular, the T-models~\cite{Kallosh:2013hoa} have minima about which the potential behaves as in Eq.~(\ref{Vk}) for even $k$. Other examples of this type \cite{Garcia:2020eof}, can be derived from no-scale supergravity.

No-scale supergravity \cite{no-scale,LN} is an ideal framework for the construction of inflationary models \cite{GL,KQ,EENOS,eno6,eno7,DLT,eno9,KLT,EGNO4,king1,enov3,building}. Indeed, it is relatively straightforward to construct
models similar to the Starobinsky model, which yield phenomenologically favorable values for the spectral tilt, $n_s$,
tensor to scalar ratio, $r$, and normalization, $A_s$ \cite{eno6,building}. 
If the inflaton is not directly coupled to the Standard Model sector, an interesting feature of these models is the complete lack of inflaton decay modes \cite{ekoty,EGNO4}. Thus, it is quite natural to have a stable inflaton and hence a dark matter candidate.

Inflation and the predictions for $n_s$, $r$, and $A_s$,
are determined by the large field values of $\phi$.
A weak scale (or SUSY-breaking) mass for the inflaton will
play no role until well after reheating is complete.
Reheating in this class of models occurs almost instantaneously
after the end of inflation, thus leading to a high reheat temperature (and a potential gravitino problem). The same interaction which leads to reheating ($\sigma \phi^2 b^2$)
also brings the inflaton into thermal equilibrium
and the inflaton relic density is determined by standard 
freeze-out conditions.  Inflaton dark matter in this respect
closely resembles a scalar singlet (Higgs portal) dark matter model \cite{Burgess:2000yq,Belanger:2013xza,Cline:2013gha,Duerr:2015mva,Abe:2014gua,noz,mnoz,noz2,hooper,oleg,Higgsportal,Arcadi:2017kky}. To avoid the constraints from direct detection experiments  \cite{XENON,LUX,PANDAX}, we must be in one of two mass regimes for $m_\phi$: either $\sigma$, is relatively large (of order 1) and the inflaton mass is in the range $m_\phi \sim 1 -  5$ TeV (where the upper limit stems from the perturbativity of the couplings) or $\sigma \sim 10^{-4} - 4 \times 10^{-3}$ and
$m_\phi \simeq m_h/2 = 62.6$ GeV, and the relic density is
determined by the resonant annihilation of the inflaton. 
For very large values of 
$\sigma$ there is a delay in reheating due to the kinematic suppression of the scattering rate (discussed below).
However, once oscillations begin, 
 reheating is essentially instantaneous and a maximal reheating temperature $T_{\rm{RH, \, Max}}$  is reached.

In what follows, we first outline in Section \ref{sec:SG} the essentials of slow-roll inflation in T-models and their construction in the context of no-scale supergravity. Our goal is a model which
satisfies the experimental determinations of $n_s$ and $A_s$ \cite{Planck}, and the upper bound on $r$ \cite{Planck,rlimit,Tristram:2020wbi}. The potential will be of the Starobinsky type, though dominated by $\phi^k$
terms at high field values,  for $k = 4, 6$. The weak scale mass term may be explicitly present or generated through radiative corrections. 
The inflaton in these models is stable and couples to other scalars in the theory. 
In section \ref{sec:RH}, we describe the reheating process in some detail. We derive analytic expressions for the reheating temperature as a function of $\sigma$ for $k=4$, and present numerical solutions for both $k = 4, 6$. Finally, in Section \ref{sec:DM} we briefly review the 
constraints on the inflaton mass in order to provide a dark matter candidate. Our conclusions are given in Section \ref{sec:concl}.

\section{Inflation and Supergravity Models}
\label{sec:SG}

\subsection{Slow-Roll Inflation and T-Models}
In this section we recall briefly the properties of slow-roll inflation and discuss the T-models of inflation \cite{Kallosh:2013hoa}, that are characterized by the inflationary potential
\begin{equation}
    \label{inf:tmodel}
    V(\phi) \; = \; \lambda M_P^4 \left[\sqrt{6} \tanh\left(\frac{\phi}{\sqrt{6}M_p} \right) \right]^k \, ,
\end{equation}
which can be easily constructed in the context of no-scale supergravity \cite{Garcia:2020eof}, and can be approximated about a minima as $V \sim \phi^k$ for even $k$. In the slow-roll approximation one commonly neglects the field acceleration term, $\ddot{\phi}$, and the scalar field equation of motion and the dynamics of the inflaton is governed by the first-order Klein-Gordon equation:
\begin{equation}
    \label{inf:kgordon}
    3 H \dot{\phi} + V'(\phi) \simeq 0 \, ,
\end{equation}
where $H = \frac{\dot{a}}{a}$ is the Hubble parameter and $V'(\phi) \equiv \frac{d V(\phi)}{d \phi}$. To ensure that the slow-roll conditions are satisfied, one may introduce the slow-roll parameters $\epsilon$ and $\eta$, which together with the potential~(\ref{inf:tmodel}) lead to
\begin{equation}
    \label{inf:epsilon}
    \epsilon \; \equiv \; \frac{1}{2} M_P^2 \left(\frac{V'(\phi)}{V(\phi)} \right)^2 \; = \; \frac{k^2}{3} \csch^2 \left(\sqrt{\frac{2}{3}} \frac{\phi}{M_P} \right) \, ,
\end{equation}
and
\begin{equation}
    \label{inf:eta}
    \eta \; \equiv \; M_P^2 \left(\frac{V''(\phi)}{V(\phi)} \right) \; = \; \frac{2}{3} k\left[k-\cosh \left(\sqrt{\frac{2}{3}} \frac{\phi}{M_{P}}\right)\right] \csch^{2}\left(\sqrt{\frac{2}{3}} \frac{\phi}{M_{P}}\right) \, ,
\end{equation}
where the slow-roll approximation is valid as long as $\epsilon \ll 1$ and $|\eta| \ll 1$, which corresponds to a flat potential. The end of inflation occurs when the condition $\ddot{a} = 0$ is satisfied, which is equivalent to the inflaton energy density $\rho_{\rm{end}} = \frac{3}{2} V(\phi_{\rm{end}})$ which occurs when $a = a_{\rm end}$. It was shown in~\cite{Garcia:2020eof,egno5} that using these conditions, one can find an approximate solution for the field value at the end of inflation, 
\begin{equation}
    \label{inf:phiend}
    \phi_{\rm{end}} \; = \; \sqrt{\frac{3}{8}} M_P \log \left[\frac{1}{2} + \frac{k}{3} \left(k + \sqrt{k^2 + 3} \right) \right] \, ,
\end{equation}
and
\beq
\rho_{\rm end} = \frac32 \lambda \left( \frac{\phi_{\rm end}^k}{M_P^{k-4}}\right) \, .
\eeq

To compute the number of e-folds between the horizon exit at the scale $k_*$ and the end of inflation at $\phi_{\rm{end}}$, the following equation can be used:
\begin{equation}
    \label{inf:efolds}
    N_* \; = \; \int_{t_*}^{t_{\rm{end}}} H dt \; \simeq \;  \frac{1}{M_P^2} \int_{\phi_{\rm{end}}}^{\phi_*} \frac{V(\phi)}{V'(\phi)} d \phi \; \simeq \; \int_{\phi_{\rm{end}}}^{\phi_*} \frac{1}{\sqrt{2\epsilon}} \frac{d \phi}{M_P} \; \simeq \; \frac{3}{2 k} \cosh \left(\sqrt{\frac{2}{3}} \frac{\phi_{*}}{M_{P}}\right)  \, ,
\end{equation}
where typical values lie in the range $N_* \in (50, 60)$.
To connect the slow-roll parameters with the principal CMB observables  \cite{Planck,rlimit,Tristram:2020wbi}, the scalar tilt, $n_s$, the tensor-to-scalar ratio, $r$, and the amplitude of the curvature power spectrum, $A_s$, one can use the following expressions,
\begin{eqnarray}
{\rm Amplitude~of~scalar~perturbations}~A_s:\; 
A_s \;& = &\; \frac{V(\phi_*)}{24 \pi^2 \epsilon_* M_P^4} \simeq 2.1 \times 10^{-9} \, , \label{eq:As} \\ \notag
\hspace{-15mm} {\rm Scalar~spectral~tilt}~n_s:\;  n_s \; & \simeq &\; 1 - 6 \epsilon_* + 2 \eta_*\\
& = &\; 0.965 \pm 0.004 \; (68\%~{\rm CL}) \, ,
\label{eq:ns} \\
\hspace{-5mm} {\rm Tensor\mbox{-}to\mbox{-}scalar~ratio}~r:\;  r \; &
\simeq & \; 16 \, \epsilon_* < 0.061 \; \,  (95\%~{\rm CL}) \, , \label{eq:r}
\label{observables}
\end{eqnarray}
where these variables are evaluated at Planck pivot scale $k_* = 0.05 \, \rm{Mpc}^{-1}$. For T-models with potential form~(\ref{inf:tmodel}), we find that the CMB observables are given by
\begin{equation}
    \label{inf:obs3}
    n_s \; \simeq \; 1 - \frac{2}{N_*} - \frac{3}{2 N_*^2}, \qquad r \; \simeq \; \frac{12}{N_*^2} \, ,
\end{equation}
and
\begin{equation}
    \label{inf:obs2}
    A_{S_*} \simeq \frac{6^{\frac{k}{2}}}{8 k^{2} \pi^{2}} \lambda \sinh ^{2}\left(\sqrt{\frac{2}{3}} \frac{\phi_{*}}{M_{P}}\right) \tanh ^{k}\left(\frac{\phi_{*}}{\sqrt{6} M_{P}}\right) \, .
\end{equation}
In order to find the normalization constant of the potential, $\lambda$, we can combine Eqs.~(\ref{inf:efolds}) and~(\ref{inf:obs2}), which leads to
\begin{equation}
    \label{inf:lambda}
    \lambda \; \simeq \; \frac{18 \pi^{2} A_{S_*}}{6^{k / 2} N_{*}^{2}} \, ,
\end{equation}
and in the range of $N_* \in (50, 60)$ this approximation  has an error of only 3\% when compared to a fully numerical evaluation. 

In Section~\ref{sec:RH}, we discuss the reheating mechanism which is achieved through the scattering process $\phi \phi \rightarrow b b$ for $k = 4, \, 6$. Because we assume that this process only occurs after the inflaton field passes the zero point for the first time at time $t_i$ when $a = a_i$, which ensures that we have at least one oscillation, we must evaluate numerically the inflaton energy density $\rho_{\phi}(t_i) \equiv \rho_i$. For the nominal choice of $N_* = 55$, we find the following parameters:
\begin{align}
    \label{inf:parameters4}
    k = 4 \, , && \lambda \simeq 3.4 \times 10^{-12} \, , && \rho_i \simeq 7.32 \times 10^{-13} \, M_P^4 \, , && a_i/a_{\rm{end}} \simeq 2.1 \, ; \\
    \label{inf:parameters6}
    k = 6 \, , && \lambda \simeq 5.7 \times 10^{-13} \, , &&  \rho_i \simeq 2.43 \times 10^{-13} \, M_P^4 \, , && a_i/a_{\rm{end}} \simeq 2.6 \, ,
\end{align}
where in both cases $n_s \simeq 0.963$ and $r \simeq 3.97 \times 10^{-3}$.

\subsection{Supergravity}
In this Section we review some general features of supergravity models and illustrate how to construct simple WIMPflation models in the no-scale supergravity framework.\footnote{For a recent review on no-scale supergravity, see~\cite{building}.} A generic supergravity theory and its geometry is characterized by the K\"ahler potential, $K(\phi^i, \phi_j^*)$, which is a function of complex chiral and anti-chiral fields. To account for particle interactions, one can introduce a holomorphic function of the chiral fields, called the superpotential, $W(\phi^i)$, and define the K\"ahler function, $G \, \equiv K \, + \ln |W|^2$. The corresponding supergravity action is given by
\begin{equation}
\label{sg:action}
 S \; = \; \int d^4 x \sqrt{-g} \left(K_{i}^{j} \partial_{\mu} \phi^{i} \partial^{\mu} \phi_{j}^{*} - V \right) \, ,
\end{equation}
where $K_{i}^{j} \equiv \partial^{2} K / \partial \phi^{i} \partial \phi_{j}^{*}$ is the K\"ahler metric, and the effective scalar potential is~\footnote{We work in Planck units with $M_P = 1/\sqrt{8 \pi G_N} = 1$.}  
\begin{equation}
    \label{sg:pot}
    V \; = \; e^G \left[\frac{\partial G}{\partial \phi^i} \left(K^{-1} \right)^{i}_{j} \frac{
    \partial G}{\partial \phi_j^*} - 3 \right] \, ,
\end{equation}
where $ \left(K^{-1} \right)^{i}_{j}$ is the inverse K\"ahler metric.

In order to construct a viable model of WIMPflation in the context of supergravity, we consider a simple toy model and introduce the following no-scale K\"ahler potential:
\begin{equation}
    \label{sg:kah}
    K \; = \; -3 \log \left(T + T^* - \frac{|\phi|^2}{3} - \frac{|H|^2}{3} - \frac{|\bar{H}|^2}{3} \right)  \, ,
\end{equation}
where the field $T$ corresponds to the volume modulus, $\phi$ is a matter-like field associated with the inflaton, and $H$ and $\bar{H}$ are a pair of Higgs multiplets 
with opposite hypercharge. 

Decays of the inflaton in no-scale supergravity models were considered in detail in \cite{ekoty,EGNO4}. Decays to matter fermions or scalars require, as a necessary but not sufficient condition, $\langle W_\phi \rangle \ne 0$. Similarly, a decay to gauginos and matter fermions or to gravitinos require
$\langle G_\phi \rangle \ne 0$. Indeed without a direct coupling to matter (for example when the inflaton is associated with the scalar partner of the right-handed neutrino \cite{eno8}), the only remaining decay channels are to pairs of gauge bosons and gauginos, and this requires a non-trivial dependence of the gauge kinetic function on $\phi$. Without this dependence,
the inflaton is naturally stable in no-scale models. 

To obtain the potential of the form $V = \lambda \phi^k$, we postulate the following superpotential,
\begin{equation}
\label{sg:sup1}
    W \; = \; \sqrt{\frac{\lambda}{3}} \phi^2 \left(\phi^2 - 3 \right)  \left( T - \frac{1}{2} \right) - c + d \, \phi^2 + \mu H \bar{H} \, ,
\end{equation}
where the first term is responsible for inflation, $\mu$ is a $\mu$-term, $c = m_{3/2}$ is the gravitino mass and associated with supersymmetry breaking, and $d$ is a constant that is necessary to ensure that we do not have a large quadratic term in the potential, which may lead to an excessively large inflaton dark matter mass. If we use the K\"ahler potential~(\ref{sg:kah}) with the superpotential~(\ref{sg:sup1}), we find from Eq.~(\ref{sg:pot}) that the potential is given by
\begin{equation}
    V \; \simeq \; \lambda \phi^4 + 2 \left(2d^2 -\sqrt{3 \lambda} c \right) \phi^2 \, ,
\end{equation}
where we have assumed that the volume modulus is stabilized and fixed dynamically to $\langle T \rangle = \frac{1}{2}$ by some higher-order stabilization terms in the K\"ahler potential~\cite{ekn3,strongpol,dlmmo,ADinf,ego,eno7,EGNO4}. To fix the quadratic term we choose $m_\phi^2 \sim 8(d^2 - \sqrt{3 \lambda} c/2)$
where we allow for the additional weak scale
contributions from the Higgs expectation value, $\sigma v^2$, as well as radiative corrections from supersymmetry breaking. In this model, at the minimum, when $\langle T \rangle = \frac{1}{2}$ and $\langle \phi \rangle = 0$, the mass of the gravitino is given $m_{3/2} \; = \; e^{K/2} |W| \; = \; |c|$, which leads to supersymmetry breaking due to non-vanishing $F$-term. Thus $d \sim \lambda^{1/4} \sqrt{m_{3/2} M_P}$.

The final term in the effective potential~(\ref{Vmk}),  $\sigma b^2 \phi^2$, with $b \equiv h$, where $h$ is the light scalar component of the chiral superfields $H$ and $\bar{H}$, can be generated by a strongly-stabilized Giudice-Masiero (GM) term~\cite{Giudice:1988yz} of the form
\beq
c_{\rm GM} \frac{|T|^2}{\Lambda^2}(H \bar{H} + H^* \bar{H}^*)
\eeq
added to the argument of the log in Eq.~(\ref{sg:kah}) with $\Lambda \ll M_P$ \cite{strongpol,dlmmo,ADinf,ego}. The inflaton-Higgs coupling
$\sigma$ is then proportional to the GM coupling $c_{\rm GM}$. The full details of such a model lies beyond the scope of this paper.

\section{Reheating}
\label{sec:RH}

Having established a plausible model for inflation
with a potential of the form in Eq.~(\ref{Vmk}), 
we are now in a position to discuss the prospects of reheating in this model. We follow closely the analysis derived recently in \cite{Garcia:2020wiy}. During and after inflation, the inflaton evolves according to its equation of motion
\beq\label{eq:phieom}
\ddot{\phi} + (3H + \Gamma_\phi) \dot{\phi} + V'(\phi) \;=\;0 \, ,
\eeq
where $\Gamma_\phi$ is the inflaton decay rate, or in this case scattering rate. To characterize the inflaton motion after inflation, we introduce the parameterization 
\begin{equation}
    \label{eq:anh}
    \phi(t) \; = \; \phi_0 (t) \cdot \mathcal{P}(t),
\end{equation}
where $\phi_0(t)$ encodes the redshift and dissipation effects of the inflaton envelope and $\mathcal{P}(t)$ is a quasi-periodic function which characterizes the anharmonic nature of the oscillation. In this case, the mass of the inflaton, $m_{\phi}(t)$, is defined by
\begin{equation}
    \label{eq:massinf}
    m_{\phi}^{2}(t) \equiv V^{\prime \prime}\left(\phi_{0}(t)\right)=k(k-1) \lambda M_{P}^{2}\left(\frac{\phi_{0}(t)}{M_{P}}\right)^{k-2} \, ,
\end{equation}
where in the second equality, we have ignored the weak scale mass, $m_\phi$. 

\begin{figure}
\centering
\includegraphics[width=0.53\textwidth]{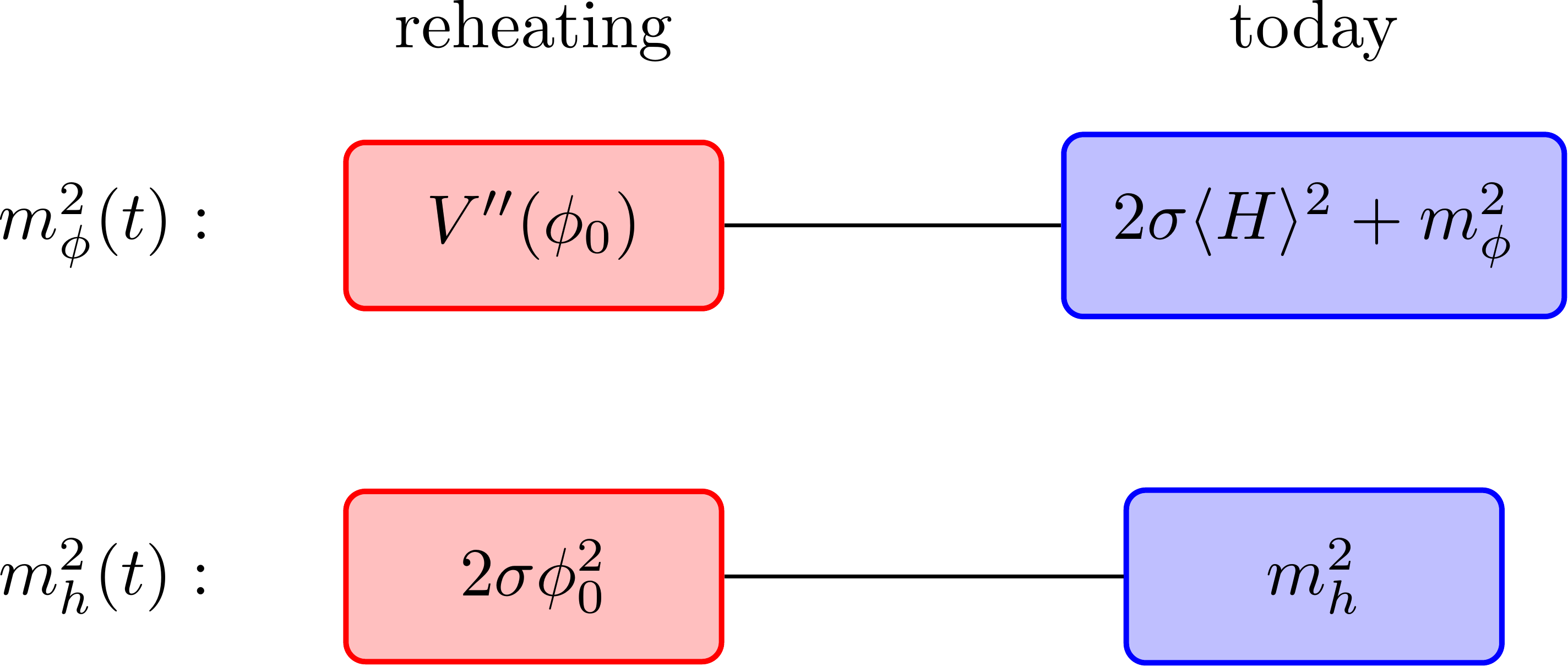}
	\caption{Instantaneous mass parameters for the inflaton and its scalar decay product at early and late times, in the case $b$ is identified with the Higgs field. During reheating $m_h\equiv m_{\rm eff}$.  } 
	\label{fig:mdiag}
\end{figure}

From the interaction term in Eq.~(\ref{Vmk}), $\sigma \phi^2 b^2$, we find that the time-dependent effective mass of the bosons is $m_{\rm{eff}}^2(t) = 2 \sigma \phi(t)^2$. It should be noted that the condition $m_{\rm{eff}}^2(t)\ll m_{\phi}^2(t)$, which ensures an efficient production of radiation from inflaton decay or scattering, is a time-dependent statement and is typically not satisfied in the models considered. The mass parameters at early and late times are shown schematically in Fig.~\ref{fig:mdiag}. To characterize this process, we average over the oscillations of $\phi$ and introduce the induced mass parameter \cite{Garcia:2020wiy},
\begin{equation}
    \label{eq:indmass}
    \mathcal{R}(t) \equiv \frac{16 \sigma}{\pi k^2 \lambda} \left(\frac{\Gamma(\frac{1}{k})}{\Gamma(\frac{1}{2} + \frac{1}{k})}\right)^2 \left(\frac{\phi_0(t)}{M_P} \right)^{4 - k} \propto \left(\frac{m_{\rm{eff}}}{m_{\phi}(t)} \right)^2  |_{\phi \rightarrow \phi_0} \, .
\end{equation}
In the models considered, $\sigma \gtrsim 10^{-4} \gg \lambda$,  $\mathcal{R} \gg 1$, and we must take into account the phase-space dependence on $m_{\rm{eff}}$. 

The loss of the inflaton energy and the production of radiation is characterized by the scattering rate \cite{Garcia:2020wiy},
\beq
\Gamma_{\phi \phi \rightarrow b b}  = \frac{\sigma_{\rm eff}^2}{8 \pi} 
\frac{\rho_\phi(t)}{ m^3_\phi(t)} =  \frac{\sigma_{\rm eff}^2 M_P}{8 \pi \left[ k(k-1) \right]^{3/2} \lambda^{3/k}} \left(\frac{\rho_\phi}{M_P^4}   \right)^{\frac3k-\frac12} \equiv \gamma_\phi \left(\frac{\rho_\phi}{M_P^4}   \right)^{\frac3k-\frac12} \, ,
\label{eq:gammappbb}
\eeq
where $\rho_\phi$ is the average energy density determined by $V(\phi_0)$. 
For $\mathcal{R} < 1$, the effective coupling, $\sigma_{\rm eff}$, is simply the Lagrangian coupling, $\sigma$. However, for $\mathcal{R} \gtrsim 1$,
the scattering rate is suppressed due to the effective mass of the final state bosons. Na\"ively, one may expect that the final state suppression is exponential, $e^{-n\pi}$, where $n$ is a Fourier mode of the inflaton, which must be taken to be very large ($n \sim \sqrt{\sigma/\lambda}$). However, once inflaton oscillations commence, during some fraction of each oscillatory period, the final state masses become small enough to allow the scattering to proceed. This results in a power-law suppression $\propto {\mathcal{R}}^{-1/2}$ rather than an exponential suppression \cite{Garcia:2020wiy}.\footnote{This suppression is present in the absence of parametric resonance~\cite{preheating, preheating2, stb, kt,Greene:1997fu}. As we argue below, this resonance will not play a role thanks to the prompt decay of $b$.}
As a result, 
the effective coupling, $\sigma_{\rm eff}$, is related to the 
potential coupling, $\sigma$, through \cite{Garcia:2020wiy}
\beq 
\sigma_{\rm{eff}}^2 (k) \; \equiv \; \frac{1}{8} k(k+2)(k-1)^2 \frac{\omega}{m_{\phi}(t)} \alpha_k \mathcal{R}(t)^{-1/2},
\eeq
where $\omega$ is the oscillation frequency. If we combine this with Eq.~(\ref{eq:indmass}), we find
\beq 
\sigma^2_{\rm eff}(k) \; = \; \frac{\pi k^{5/2}}{32\sqrt{2}} (k+2) (k-1)^{3/2} \lambda^{1/2} \left( \frac{\Gamma(\frac{1}{2}+\frac{1}{k})}{\Gamma(\frac{1}{k})} \right)^2 \left(\frac{\phi_0(t)}{M_P} \right)^{\frac{k-4}{2}} \alpha_k \sigma^{3/2}\,,
\label{Eq:sigmaeff}
\eeq
which accounts for the suppression for producing bosons with effective
masses $m_{\rm{eff}}^2 (t) = 2 \sigma \phi^2(t)$ due to the condensate, $\phi$. From the numerical evaluations, we find that the coefficient $\alpha_k = 1.22(1.35)$ for $k = 4(6)$. Roughly, suppression occurs when 
$\sigma \gtrsim \lambda\left(\frac{\phi_0}{M_P}\right)^{k-4}$ or $\mathcal{R}(t) \gtrsim 1$, which will always be the case in the parameter space required for dark matter constraints, where $\sigma \gtrsim 10^{-4}$.
Thus, we can write
\beq \label{eq:gammaphis}
\gamma_\phi = \frac{k (k+2)}{256\sqrt{2}}\left( \frac{\Gamma(\frac{1}{2}+\frac{1}{k})}{\Gamma(\frac{1}{k})} \right)^2 \left(\frac{\phi_0(t)}{M_P} \right)^{\frac{k-4}{2}} \lambda^{\frac12-\frac3k} \alpha_k \sigma^{3/2} {M_P}\simeq
\begin{cases}
0.0092 \lambda^{-1/4} \sigma^{3/2} {M_P}& k=4 \, ,\\
0.0106 \left(\frac{\phi_0(t)}{M_P}  \right) \sigma^{3/2} {M_P}& k = 6 \, .
\end{cases}
\eeq

The equation of motion~(\ref{eq:phieom}) can be rewritten in terms of the energy density and pressure stored in the scalar field,
\beq
\rho_\phi = \frac{1}{2} \dot \phi^2 + V(\phi);
~~~P_{\phi} = \frac{1}{2} \dot \phi^2 - V(\phi) \, ,
\label{Eq:rhophi0}
\eeq
as
\beq
\dot \rho_\phi + (3 H + \Gamma_\phi) (1+ w_\phi) \rho_\phi  = 0 \, ,
\label{Eq:conservation}
\eeq
where 
$w_\phi = P_\phi/\rho_\phi = (k-2)/(k+2)$.
This last equation can be expressed as
\beq
\frac{d\rho_{\phi}}{da} = -\frac{3}{a}  \frac{2k}{k+2} \left(1+ Q \right)\rho_\phi \, ,
\label{Eq:conservation2}
\eeq
where $Q \equiv \Gamma_{\phi}/3H$. When inflation ends at $a = a_{\rm end}$, defined when $\phi = \phi_{\rm end}$, the radiation energy density (composed of scalars $b$ which we assume thermalize rapidly) begins to grow and evolves according to 
\beq
\frac{a}{3}\frac{d\rho_R}{da} + \frac{4}{3}\rho_R \;\simeq\; (1+w_{\phi})Q(t)\rho_{\phi}\, ,
\label{eq:aeom}
\eeq
which together with the Friedmann equation
\beq
\rho_{\phi}+\rho_{R} \;=\; 3H^2 M_P^2\,,
\label{hub}
\eeq 
allows one to solve for $\rho_\phi(t), \rho_R(t)$, and $a(t)$ simultaneously and effectively for $\rho_\phi(a)$
and $\rho_R(a)$.

Approximate analytic solutions to Eqs.~(\ref{Eq:conservation}) - (\ref{hub}) depend on the relative sizes of $\Gamma_\phi$ and $H$.
From 
\beq
Q  \; = \; \frac{\Gamma_\phi}{3H}= \frac{\alpha_k \sigma^{3/2}}{256 \sqrt{6}} \frac{k(k+2)}{\sqrt{\lambda}}
\left(\frac{\Gamma(\frac{1}{k}+\frac{1}{2})}{\Gamma(\frac{1}{k})}\right)^2
\left(\frac{\phi_0(t)}{M_P}\right)^{\frac{2-k}{2}}
\label{Eq:ratio}
\eeq
we see that the scattering rate dominates the friction term at the end of inflation 
for $\sigma \gtrsim 10^{-3} \left(\frac{\lambda}{10^{-12}}\right)^{1/3}\left(\frac{\phi_{\rm end}}{M_P}\right)^{\frac{k-2}{3}}$.
Notice also that we recover the fact that, for $k=2$, $Q$ is constant, and reheating is not achievable by pure scattering.

For sufficiently small $\sigma$, $Q < 1$,
and can be neglected in Eq.~(\ref{Eq:conservation2}).
These solutions were derived in \cite{Garcia:2020wiy}, and we quote those solutions here:
\beq
\rho_\phi(a) = \rho_{\rm end} \left(\frac{a_{\rm end}}{a} \right)^{\frac{6k}{k+2}} \, ,
\label{Eq:rhophi}
\eeq
valid at early times when $Q \ll 1$;
and
\beq
\rho_R \;=\; \begin{cases}
\dfrac{\sqrt{3}k}{2k-5} \gamma_\phi \rho_{\rm end}^{3/k} M_P^{3-\frac{12}{k}} \left( \dfrac{a_{\rm end}}{a} \right)^4
\left[ \left( \dfrac{a}{a_{\rm end}}\right)^{\frac{4k-10}{k+2}} -1\right]\,, & \mathcal{R}\ll1\,,\\[13pt]
\dfrac{2\sqrt{3}k}{k+2} \gamma_\phi \rho_{\rm end}^{3/k} M_P^{3-\frac{12}{k}} \left( \dfrac{a_{\rm end}}{a} \right)^{\frac{k+20}{k+2}}
\left[ \left( \dfrac{a}{a_{\rm end}}\right) -1\right]\,, & \mathcal{R}\gg1\,.
\end{cases}
\label{Eq:rhoR}
\eeq
Note that for $k=4$ both expressions are one and the same.

For relatively small coupling, $\sigma$, the radiation density grows until the temperature reaches a maximum.  From Eq.~(\ref{Eq:rhoR}) we can compute the maximum of $\rho_R$ which occurs when
\beq
a_{\rm max}= a_{\rm end}\left(\frac{2k+4}{9} \right)^{\frac{k+2}{4k-10}} \, .
\label{maxend}
\eeq
However, for $a_{\rm max} < a_i$, where $a_i$ is given in Eqs.~(\ref{inf:parameters4}) and (\ref{inf:parameters6}), there is a delay in the growth of the radiation bath as the first oscillation begins. Note that $a_i$ is independent of $\sigma$. Thus the maximum temperature attained is $T_{\rm RH} \propto \rho_i^{1/4}$. 
Subsequently, the temperature, given by $T \propto \rho_R^{1/4}$, falls as $T \propto a^{\frac{9}{2k+4}}$. Note that Eq.~(\ref{Eq:rhoR})
is only valid for $2k>5$. For $k = 2$, the solution for $\rho_R$ at late times scales as
$a^{-4}$. As a result, annihilations can not lead to reheating as $\rho_\phi$ scales as $a^{-3}$ and always dominates over the radiation density.

We define the reheat temperature when 
$\rho_R = \rho_\phi$. For $k=4$, we can find
an analytical solution for $a_{\rm RH}/a_{\rm end}$ (valid for $Q \ll 1$ and $\mathcal{R} \gg 1$):
\beq
\left(\frac{a_{\rm RH}}{a_{\rm end}} - 1 \right) = \frac{\sqrt{3} \rho_{\rm end}^{1/4}}{4 \gamma_\phi} \, .
\eeq
This leads to 
\beq
T_{\rm RH} \; = \; \frac{ (3/2)^{1/4}\lambda^{1/4} \phi_{\rm end} \sigma^{3/2}}{\alpha^{1/4} \left(  (3/2)^{1/4} A \lambda^{1/2} (\phi_{\rm end}/M_P) + \sigma^{3/2} \right)} \, ,
\label{trhapp}
  \eeq
where $\alpha = g_{\rm RH} \pi^2/30$, and $g_{\rm RH} = 915/4$ is the number of relativistic degrees of freedom at $T_{\rm RH}$ (assuming a minimal supersymmetric model spectrum) and $A = \sqrt{3}\sigma^{3/2} M_P/4 \gamma_\phi \lambda^{1/4} \simeq 47 $.

However, these expressions are not valid when $\sigma \gtrsim 10^{-4} - 4 \times 10^{-3}$ which leads to $Q \gtrsim 1$. Because reheating would occur before the inflaton field completes a single oscillation the description which averages over oscillations becomes inaccurate. We find that when $\sigma > 8 \times 10^{-4} (2 \times 10^{-4})$ for $k = 4(6)$ we no longer have a complete oscillation.

To ensure that our approximations are valid, we assume that the radiation production through scattering begins when the inflaton crosses the origin, at time $t_i$, with the energy density $\rho_i$ and the scale factor ratio $a_i/a_{\rm{end}}$ given by Eqs.~(\ref{inf:parameters4}) and~(\ref{inf:parameters6}) for $k = 4, 6$ cases, respectively.
This does not affect the general description for small values of $\sigma$ but is more accurate when $\sigma > 10^{-4}$, necessary for inflaton scalar singlet models of dark matter.

For large values of $\sigma > 10^{-4}$, and hence large $Q \gg 1$,  Eq.~(\ref{Eq:conservation2}) can be approximated by 
\beq
\frac{d \rho_\phi}{da} \; \simeq \; -\left(\frac{6k}{k+2}\right) Q \cdot
\frac{\rho_\phi}{a}
\label{Eq:rhophibis}
\eeq
Using Eqs.(\ref{Eq:ratio}), Eq.~(\ref{Eq:rhophibis}) can be solved for $k>2$
(the solution for $k=2$ is an exponential decay up to $\rho_{\phi}\simeq \rho_R$, but not beyond it)
\beq
\rho_\phi(a)=M_P^4\left[\left(\frac{\rho_i}{M_P^4}\right)^{\frac{k-2}{2k}}
-\frac{\sqrt{6}(k-2)k \alpha_k \sigma^{\frac{3}{2}}}{512 \lambda^{\frac{1}{k}}} 
\left(\frac{\Gamma(\frac{1}{k}+\frac{1}{2})}{\Gamma(\frac{1}{k})}\right)^2\ln \frac{a}{a_i}\right]^{\frac{2k}{k-2}} \, .
\eeq

Considering $\phi_0 \simeq \rm{const.}$ at the very end of inflation, this corresponds to  quasi-instantaneous reheating, where we can justify that the radiation density at reheating for large values of $\sigma$ is

\beq
\rho_{\rm RH} \; = \; \rho_R(a_{RH}=a_i) \; \simeq \; \frac{\rho_{i}}{2} \simeq 
\begin{cases}
1.2 \times 10^{61} \, \, \rm{GeV}^4 & k=4 \, , \\
4.0 \times 10^{60} \, \, \rm{GeV}^4 & k = 6 \, .
\end{cases}
\eeq
This corresponds to a reheating temperature
\beq
\label{rehtemps}
T_{\rm {RH, \, Max}} =  \alpha^{-1/4}\rho_{\rm RH}^{1/4} \simeq \begin{cases}
6.3 \times 10^{14} \, \, \rm{GeV} & k=4  \, , \\
4.8 \times 10^{14} \, \, \rm{GeV} & k = 6 \, .
\end{cases}
\eeq

\begin{figure}[!pt] 
\centering
    \includegraphics[width=0.80\textwidth]{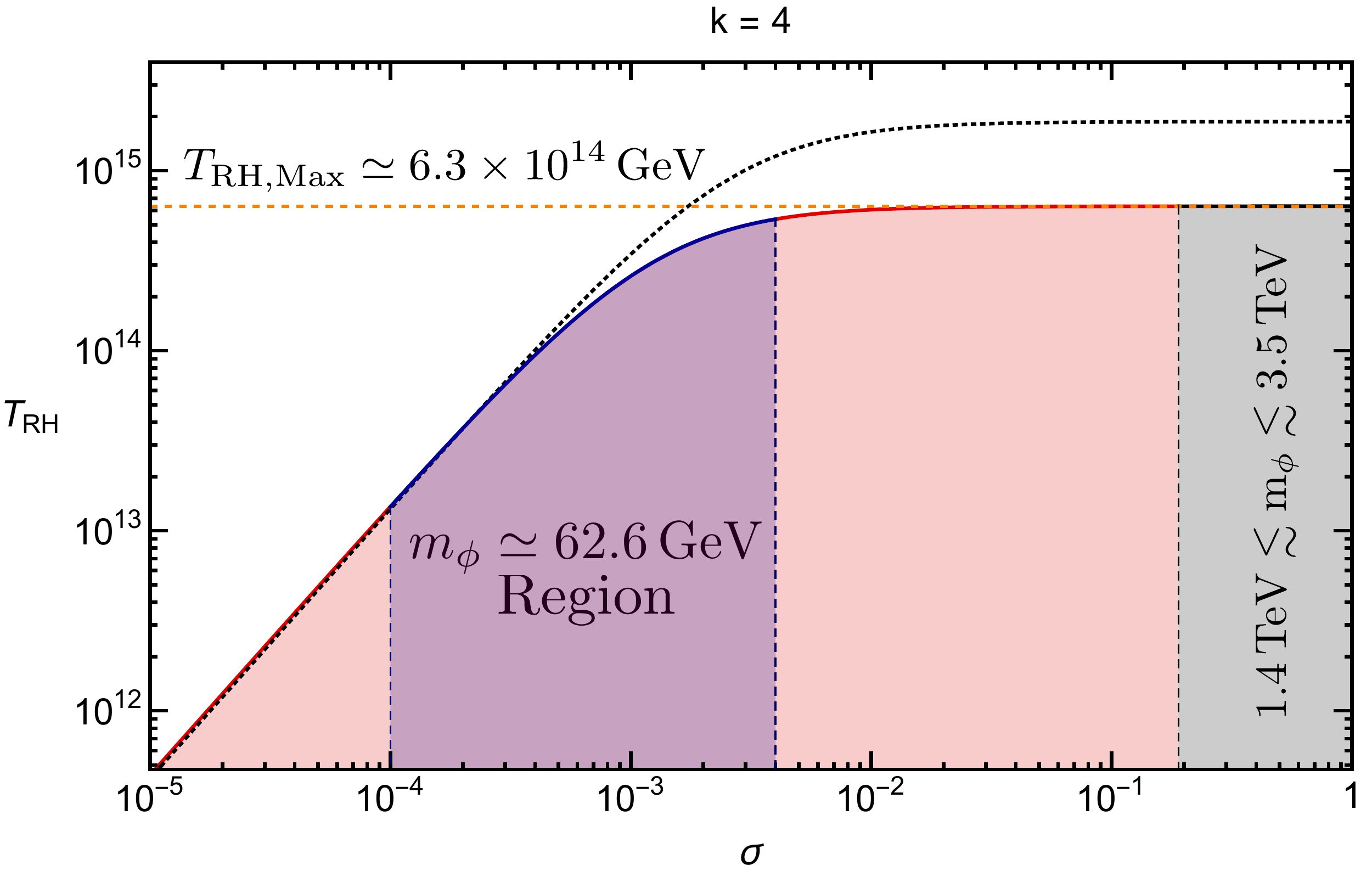}
    \includegraphics[width=0.80\textwidth]{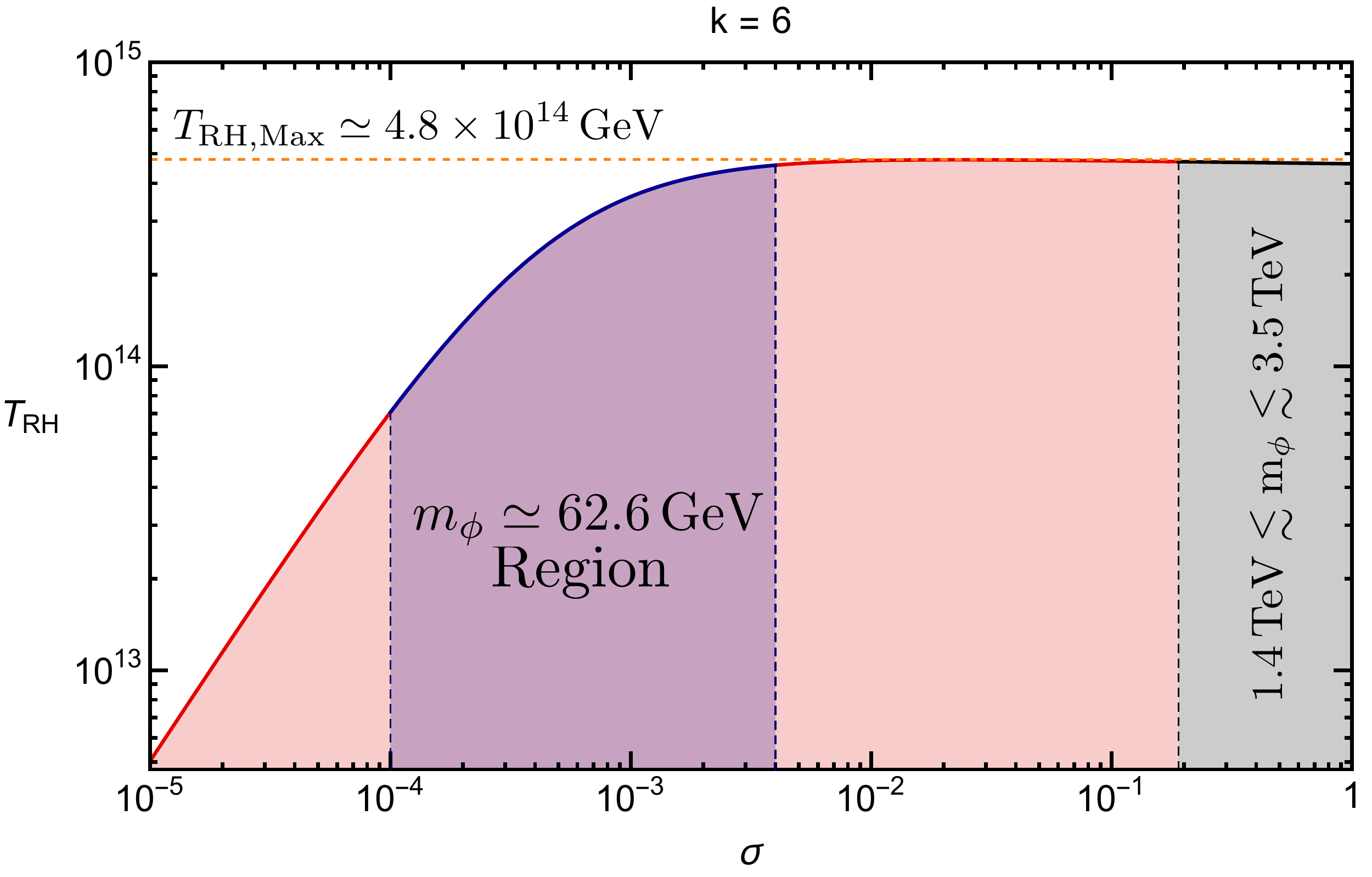}
    \caption{\textit{Plots of the reheating temperature $T_{\rm{RH}}$ as a function of coupling $\sigma$ for $k = 4$ (top) and $k = 6$ (bottom). The blue shaded region corresponds to the resonant dark matter solution, $m_{\phi} \simeq m_{h}/2 \simeq 62.6 \, \rm{GeV}$, with $10^{-4} < \sigma < 4 \times 10^{-3}$, and the gray shaded region corresponds to $m_{\phi} \gtrsim 1.4 \, \rm{TeV}$ with $\sigma \gtrsim 0.19$. For large values of $\sigma$, the reheating temperature approaches the maximum reheating temperature $T_{\rm{RH, \, Max}}$, given by Eq.~(\ref{rehtemps}). The black dotted curve for $k=4$ corresponds to the analytical approximation given in Eq.~(\ref{trhapp}).} 
    }
    \label{fig:trhvssigma}
    \vspace{-20pt}
\end{figure}

\begin{figure}[!tp]
  \centering
  \includegraphics[width=.43\textwidth]{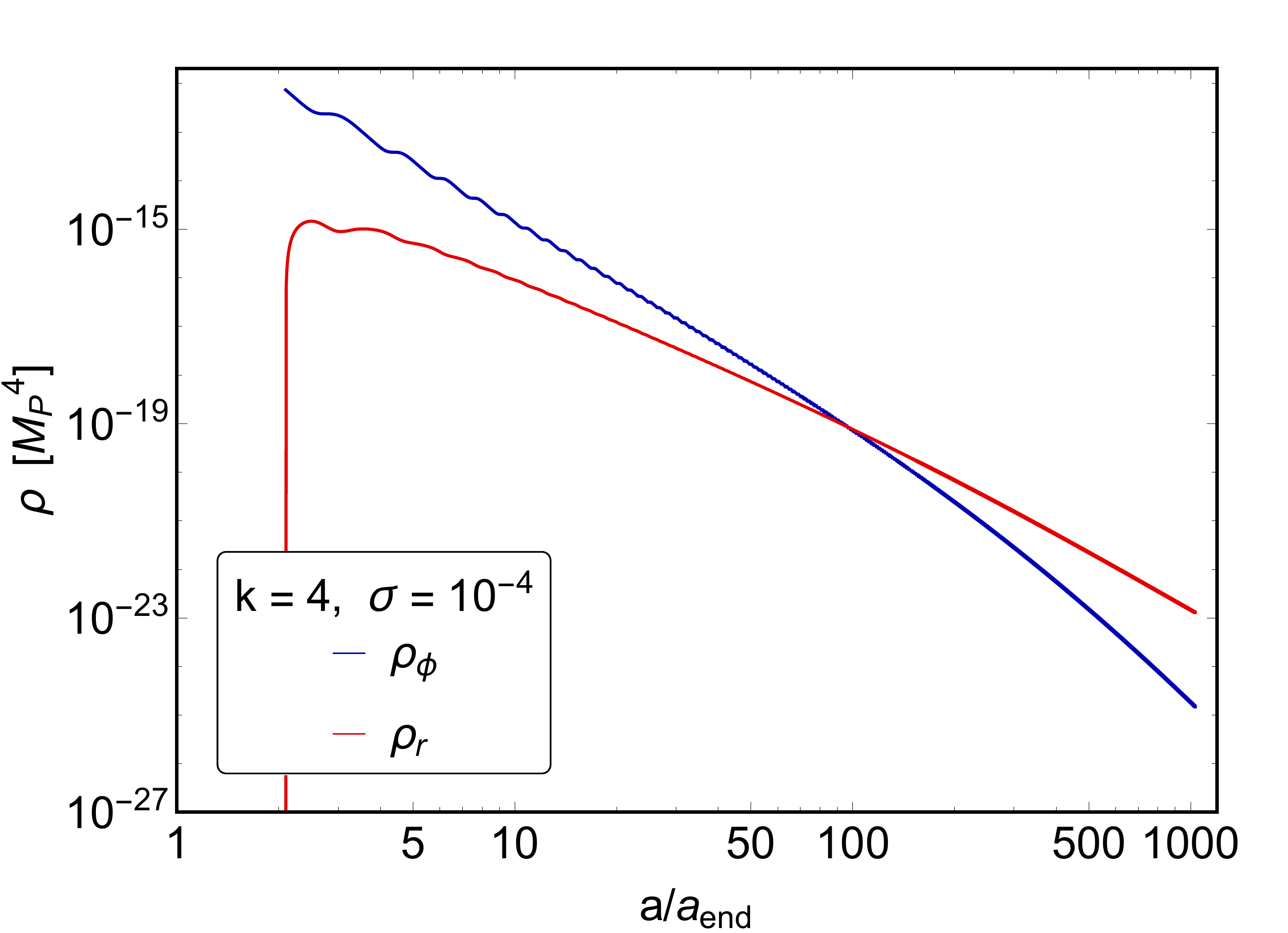}
  \includegraphics[width=.43\textwidth]{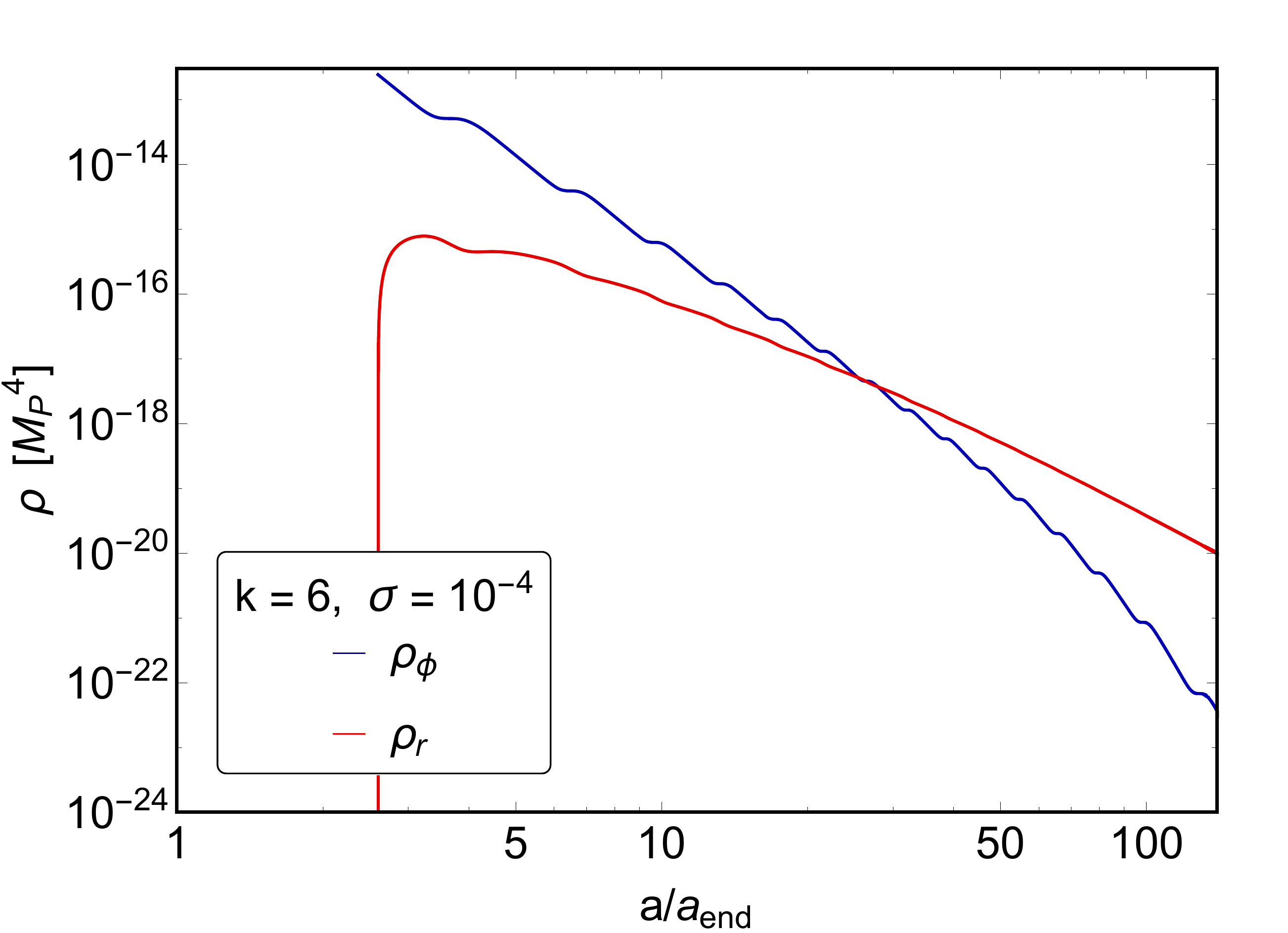}

  \vspace{0.1cm}

  \includegraphics[width=.43\textwidth]{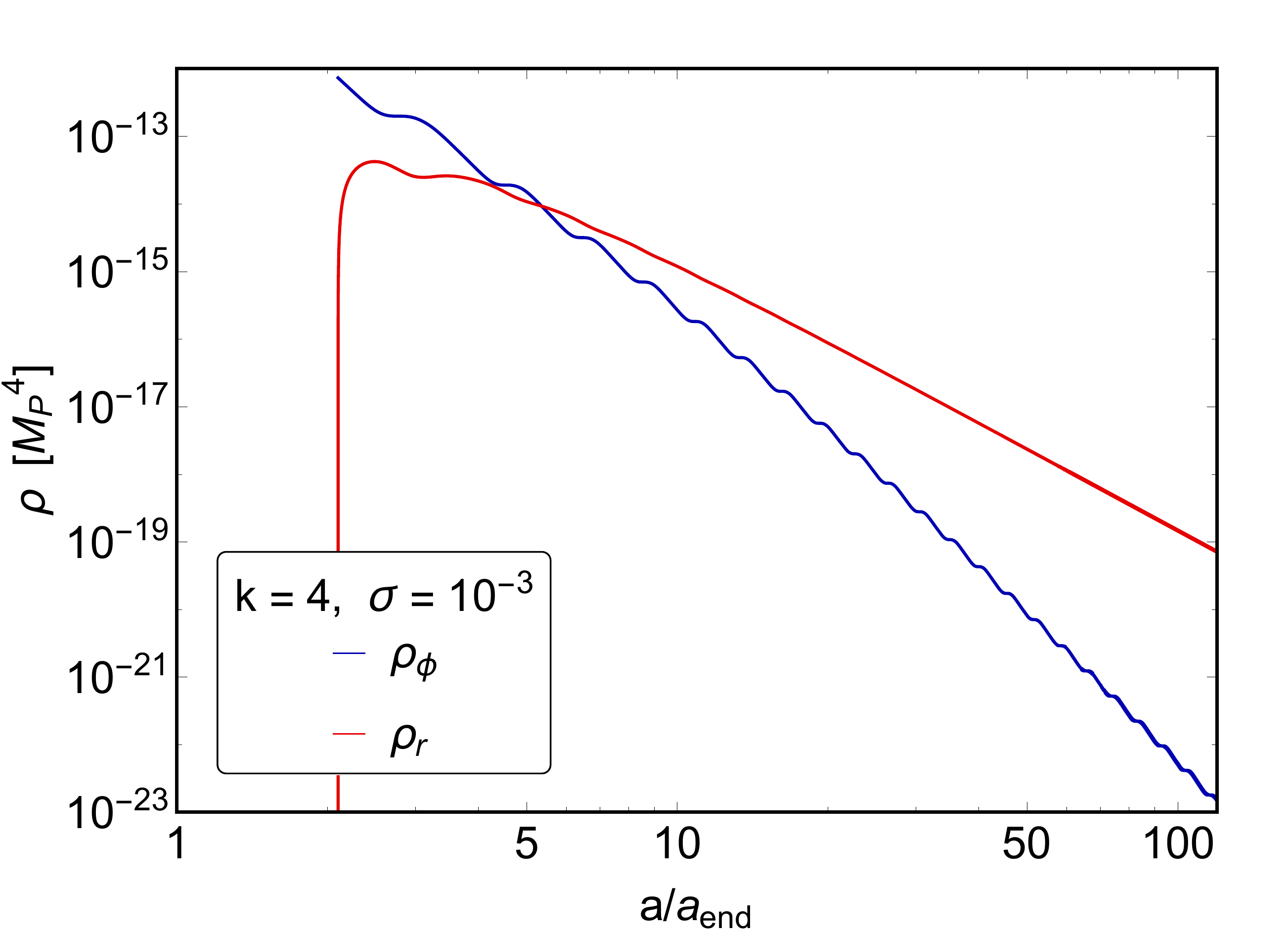}
  \includegraphics[width=.43\textwidth]{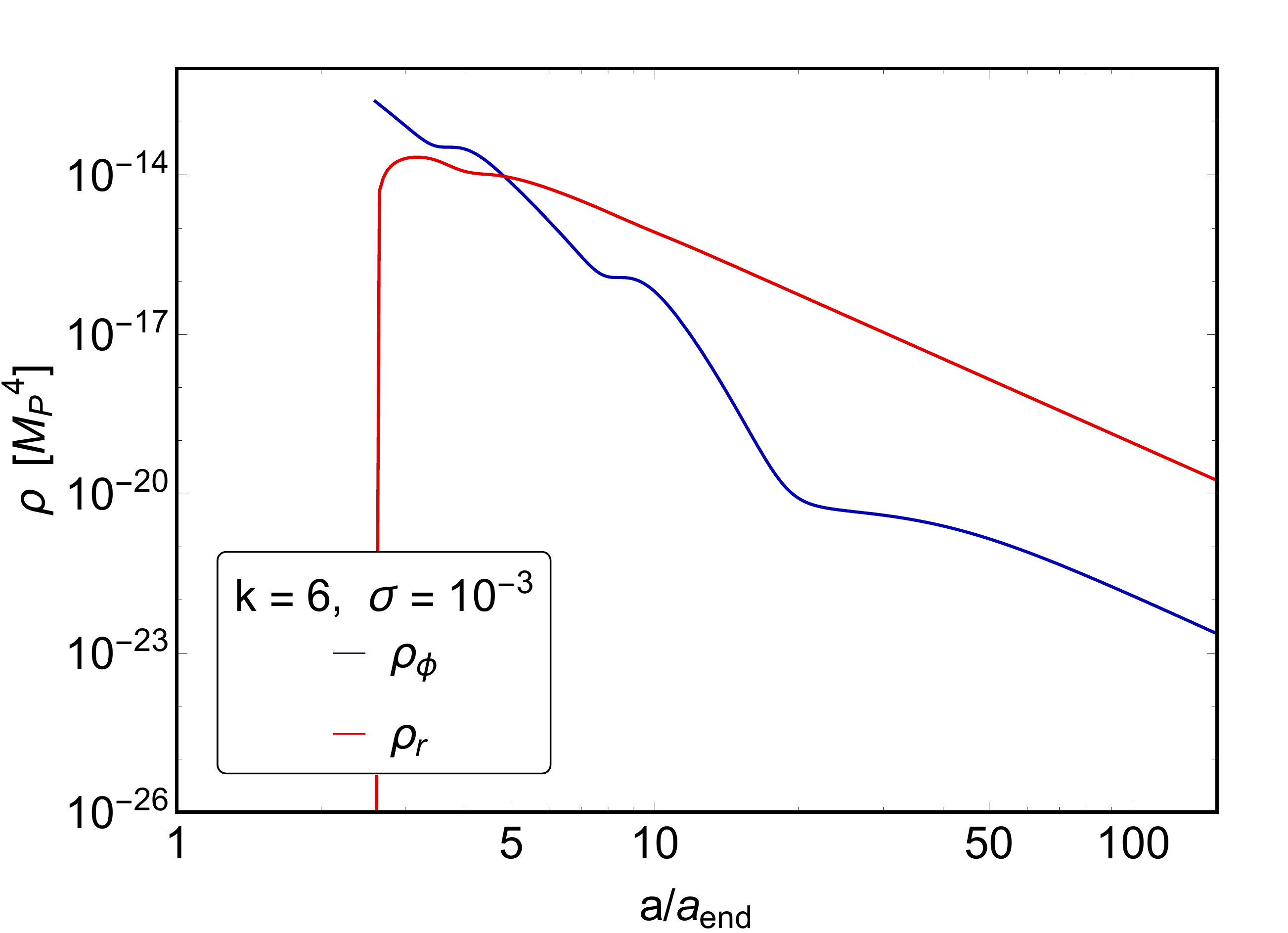}

  \vspace{0.1cm}

  \includegraphics[width=.43\textwidth]{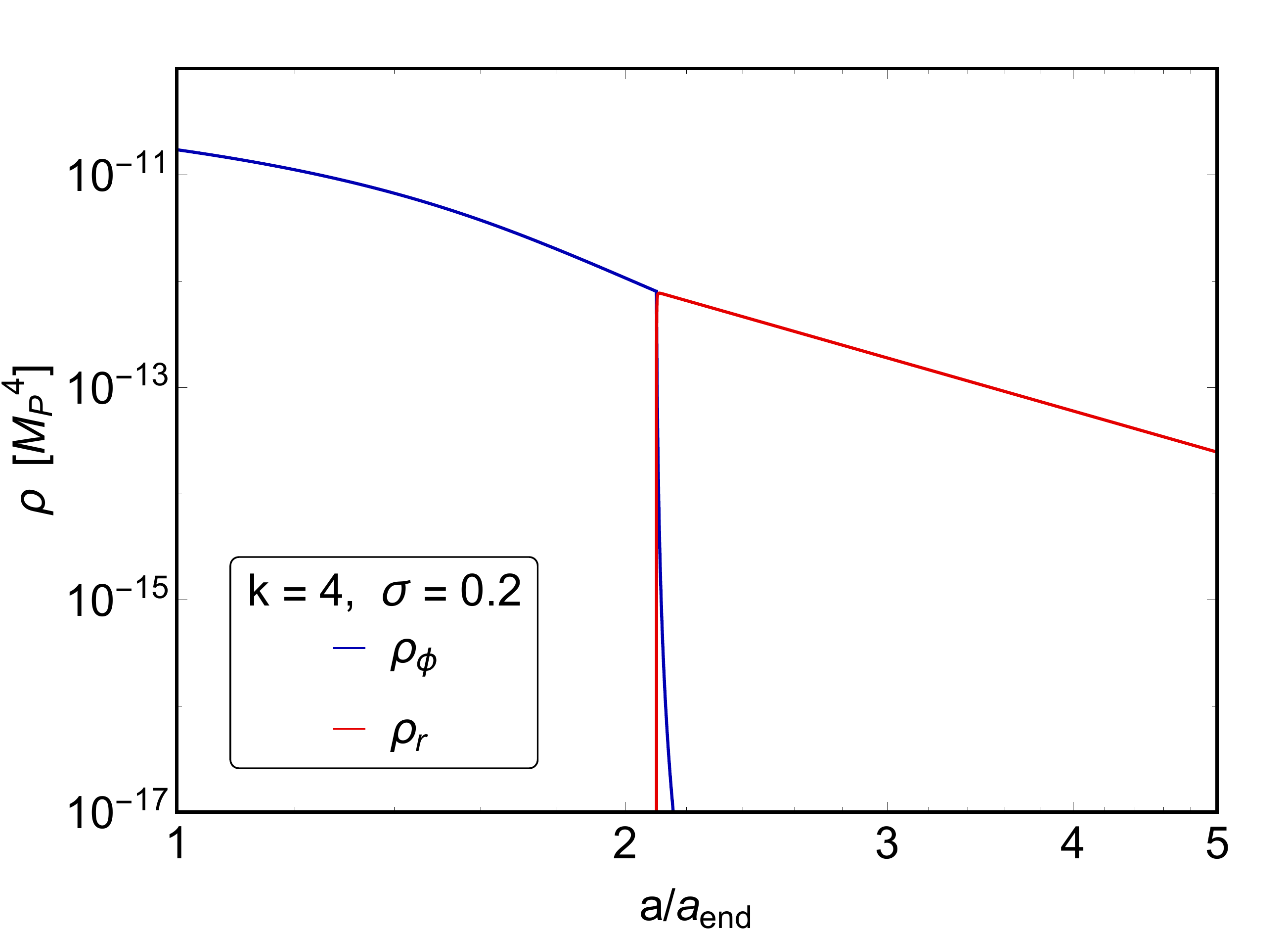}
  \includegraphics[width=.43\textwidth]{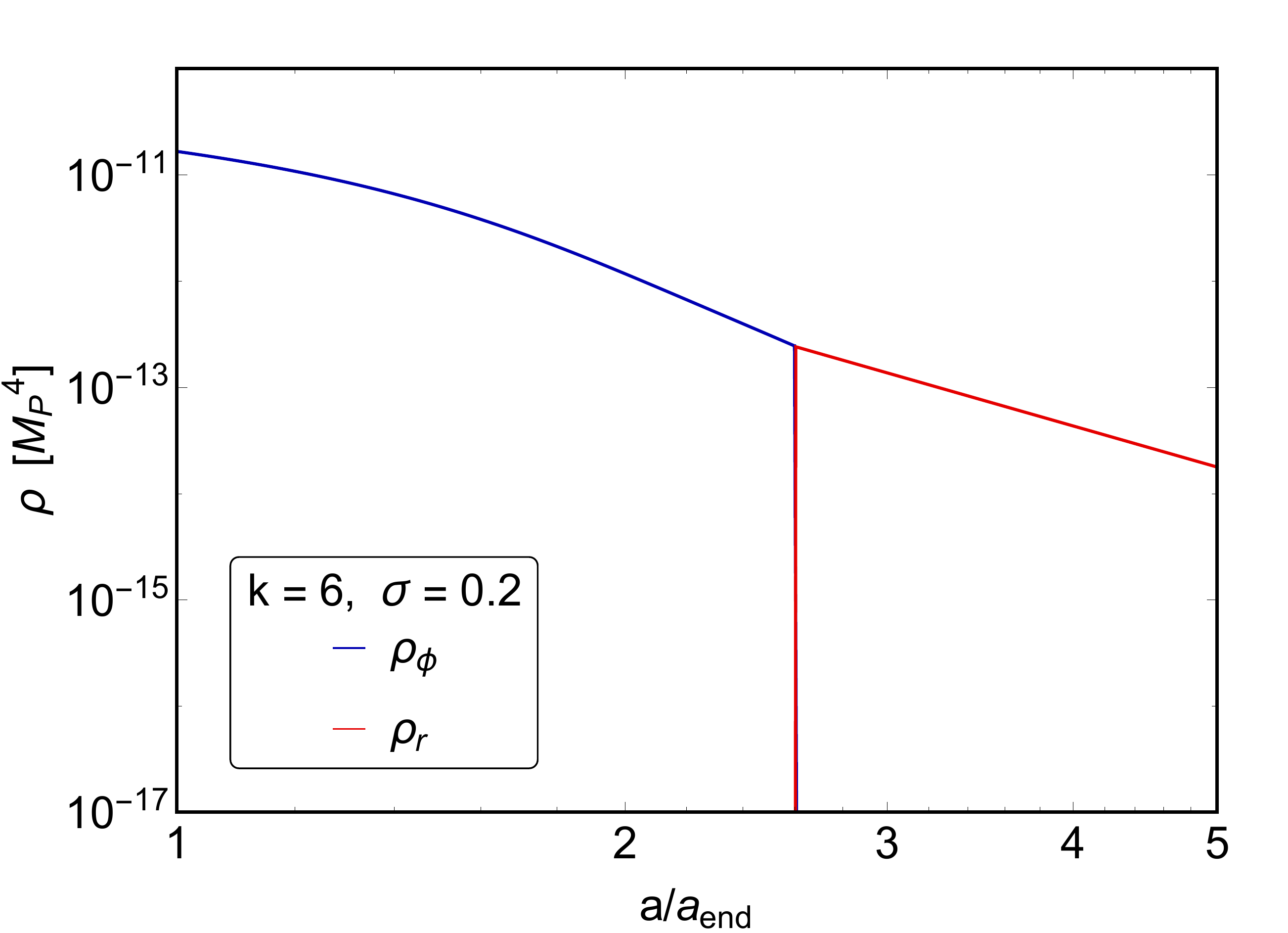}

  \caption{\it Evolution of the energy densities $\rho_{\phi}$ (blue) and $\rho_R$ (red) as a function of the scale factor ratio, $a/a_{\rm{end}}$. The left (right) panel corresponds to the case $k = 4(6)$ with values $\sigma = 10^{-4}, \, 10^{-3}$ (resonant dark matter production region) and $\sigma = 0.2$ ($m_{\phi} \gtrsim 1 \, \rm{TeV}$ region). The production of radiation through scattering begins at $a_i/a_{end} \simeq 2.1 (2.6)$ for $k = 4(6)$. As expected, for large values of $\sigma$ reheating is almost instantaneous, and this process leads to a relatively high reheat temperature of $\sim \mathcal{O} (10^{14}) \, \rm{GeV}$.}
  \label{fig:rhophi}
\end{figure}

We illustrate in Fig.~\ref{fig:trhvssigma} the values of reheating temperature $T_{\rm{RH}}$ as a function of $\sigma$ for $k = 4 \, , 6$. As expected, for large values of $\sigma$, reheating temperature approaches the maximum possible reheating temperature $T_{\rm{RH, \, Max}}$, given by Eq.~(\ref{rehtemps}). We also show the analytical approximation given in Eq.~(\ref{trhapp}) for $k=4$. The validity of the approximation clearly breaks down when $\sigma \gtrsim 10^{-3}$, as expected.
In Fig.~\ref{fig:rhophi} we show the inflaton and radiation energy densities before and after reheating for the resonant dark matter production region with $m_{\phi} \simeq m_h/2 \simeq 62.6 \, \rm{GeV}$, which corresponds to $\sigma \sim 10^{-4} - 4 \times 10^{-3}$, and $m_{\phi} = 1.4 - 5$ TeV region, which corresponds to $\sigma = 0.19 - 1$. See Section \ref{sec:DM} for more detail. It should be noted that for large values of $\sigma$ all of the inflaton energy density $\rho_i$ is converted quasi-instantaneously to radiation, which leads to a maximal temperature, $T_{\rm{Max}} = \alpha^{-1/4} \rho_i^{1/4}$. See the bottom panels with $\sigma = 0.2$ in Fig.~\ref{fig:rhophi}.

Once the radiation bath is produced, interactions between Higgs particles and inflatons, $\phi \phi \leftrightarrow h h$ at high $T$ are determined by the cross section, $\sigma v \simeq \sigma^2/4\pi E^2 \sim \sigma^2/4\pi T^2$.\footnote{Note that this differs by a factor of 2 from the cross section for the condensate $\phi$ which requires an average over oscillations \cite{Garcia:2020wiy}.} The rate for these interactions is then roughly $\Gamma_{\phi H} \sim \sigma^2 T/4 \pi$ and 
will be in equilibrium for temperatures $T < \sigma^2 M_P/4 \pi$. Equilibrium is then maintained until thermal freeze-out occurs as is described in the next section.

To ensure that the resonant boson production via the coupling $\sigma \phi^2 b^2$ does not occur, the Higgs bosons must decay faster than they are produced through parametric resonance~\cite{preheating, preheating2, stb, kt,Greene:1997fu}. The Higgs decay rate is proportional to its effective mass, $\Gamma_h \sim m_{\rm{eff}}(t)$, where $m_{\rm{eff}}(t) \simeq \sqrt{2 \sigma} |\phi(t)| $, and for large values of $\sigma$ the Higgs bosons decay instantly and resonant production does not arise. It was argued in~\cite{oleg} that for $k = 4$ case, the Higgs bosons decay faster than they are generated through parametric resonance when $\sigma/\lambda \gtrsim 10^{3}$. Since resonant dark matter production occurs when $\sigma/\lambda \gtrsim 3 \times 10^{7} \gg 10^{3}$, the limit is easily satisfied. Similar arguments can be used for $k = 6$ case.

\section {Dark Matter}
\label{sec:DM}

Inspired by the example of a massive neutrino, the most natural dark matter candidate is a Weakly Interacting Massive Particle (WIMP). 
The physics of such a particle benefits from 
the fact that the weak interaction has the advantage of being strong enough ($g_2\simeq 0.65$) to allow dark matter to annihilate efficiently when decoupling from the thermal bath, thus avoiding an over-density in violation of data from the PLANCK satellite \cite{Planck}. In addition, such a coupling has the advantage of opening up the possibility of direct detection modes via the dark matter-nucleus interaction in underground detectors such as XENON1T \cite{XENON}, LUX/LZ \cite{LUX} or PANDAX \cite{PANDAX}. These large scale experiments have 
reached the impressive limit $\sigma_{dm-p}\lesssim 10^{-46}~ \rm{cm^2}$ for a 100 GeV dark matter.  Translated into couplings, for an $s$-wave process, this corresponds to a limit $g_{dm} \lesssim 10^{-4}$, much less that the vanilla SM weak coupling.

The case of an inflaton as dark matter is completely indistinguishable from the generic model called "scalar singlet Higgs portal" \cite{Burgess:2000yq,Belanger:2013xza,Cline:2013gha,Duerr:2015mva,Abe:2014gua,noz,mnoz,noz2,hooper,oleg,Higgsportal} (for a review of models, see \cite{Arcadi:2017kky}). Indeed, the potential described in Eq.(\ref{Vmk}), with $b \equiv h$ is characteristic of this type of construction, and has been the subject of numerous studies for several years. The origin of the dark matter is rarely specified, and 
quartic couplings of the type $\sigma \phi^2 h^2$ are usually introduced because they can be introduced. In this case, $\sigma$ is a free parameter, without any relation to inflation and/or reheating.
Once we suppose that the inflaton field $\phi$ is stable, it plays the role of
the dark matter candidate once the reheating is completed.
It becomes then necessary to check if it does not violate constraints from PLANCK data, direct detection 
analysis, and Higgs lifetime measurements at LHC.

At late times, the condensate has thermalized, and the inflaton is part of thermal bath with a well defined mass, $m_\phi$. The Boltzmann equation
\beq
\frac{d n_\phi}{dt} + 3 H n_\phi 
= - n_\phi^2 \langle \sigma v \rangle_{\phi \phi\rightarrow SM} \, ,
\eeq
where $n_\phi=\frac{\rho_\phi}{m_\phi}$ can be simplified to
\beq
\frac{dY_\phi}{dT} = Y_\phi^2 \sqrt{\frac{3}{\alpha}} \langle\sigma v \rangle M_P~~~
\Rightarrow~~~ Y_\phi(T) = \sqrt{\frac{\alpha}{3}}\frac{1}{\langle \sigma v \rangle T_{FO} M_P}
\eeq
with $Y_\phi=\frac{n_\phi}{T^3}$
and $\alpha =\frac{g_{FO} \pi^2}{30}$, $g_{FO}$ being the number of degrees of freedom at the 
freeze-out temperature $T_{FO} \simeq \frac{m_\phi}{30}$.
Solving the Boltzmann equation gives 
\bea
\Omega_\phi h^2 & = & 1.6 \times 10^{11}~
Y_\phi(T_0)\left( \frac{m_\phi}{1~\rm{TeV}}\right) 
=1.6 \times 10^{11}\left(\frac{g_0}{g_{FO}} \right)
Y_\phi(T_{FO})
\left( \frac{m_\phi}{1~\rm{TeV}}\right)
\nonumber
\\
&
= &
1.6 \times 10^{11}\left(\frac{g_0}{g_{FO}} \right)
\sqrt{\frac{\alpha}{3}}\frac{30}{\langle \sigma v \rangle M_P m_\phi}
\left( \frac{m_\phi}{1~\rm{TeV}}\right) \, ,
\label{Eq:omega}
\eea
where we used number conservation, so that  $n_\phi(T_0)=
n_\phi(T_{FO})\frac{T_0^3}{T_{FO}^3}\frac{g_0}{g_{FO}}$. The relic abundance depends strongly on $\langle\sigma v\rangle$ which, in turn, depends on
the regime we are considering. Indeed, for dark mater masses above the electroweak symmetry breaking scale, the dominant annihilation channel will be
$\phi \phi \rightarrow hh$ given by the quartic interaction $\sigma \phi^2 h^2$, 
\beq
\langle \sigma v \rangle_{hh}= \frac{\sigma^2}{4 \pi m_\phi^2}
\eeq
which gives
\beq
\frac{\Omega_\phi h^2}{0.1}\simeq 
\left(\frac{0.19}{\sigma}\right)^2 \left( \frac{m_\phi}{1~\rm{TeV}}\right)^2,
\eeq
where we have used $g_0 = 3.91$, $g_{FO} = 345/4$
appropriate for $T_{FO} = m_\phi/30$ with $m_\phi = 1$ TeV.

However, we can derive another constraint on $\sigma$ (and hence $m_\phi$ from direct detection experiments). 
Indeed, the proton-inflaton scattering cross-section 
\bea
\sigma_{\phi p}&=&\frac{\sigma^2}{\pi m_h^4} \frac{m_p^4 f_p^2}{(m_\phi + m_p)^2}\simeq
9.1\times 10^{-19} \left(\frac{\sigma}{0.01}\right)^2 
\left(\frac{100~\rm{GeV}}{m_\phi} \right)^2 \rm{GeV^{-2}}
\nonumber
\\
&
\simeq &
3.5 \times 10^{-46} \left(\frac{\sigma}{0.01} \right)^2
\left(\frac{100~\rm{GeV}}{m_\phi} \right)^2~{\rm cm^2} =
1.2 \times 10^{-45} \left(\frac{0.1}{\Omega_\phi h^2}\right)~\rm{cm^2} \, ,
\label{Eq:sigmaphip}
\eea
where we took $f_p \simeq 0.3$, structure function representing the Higgs-nucleon coupling. The current limit from XENON1T is satisfied for this cross-section for masses $m_\phi \gtrsim 1.4$ TeV, opening a  window that is still allowed by the combination relic abundance and direct detection constraint. At still larger values of the coupling, there is a (supersymmetry breaking) model dependent limit on
$\sigma$ from the requirement of perturbativity of the couplings.
We do not attempt here to calculate this limit, and simply take $\sigma < \mathcal{O}(1)$ corresponding to $m_\phi \lesssim 5$ TeV.

This region is represented by the gray region on the right part of Fig.~\ref{fig:trhvssigma}. The relic abundance and direct detection constraints, in the large mass regime, impose a lower bound on $\sigma \gtrsim 0.19$ corresponding to $m_\phi > 1.4$ TeV.  The model dependent upper, from the perturbativity of $\sigma$ is  shown as $\sigma \lesssim 1$. leads to an upper bound on the inflaton mass, $m_\phi \lesssim 5$ TeV.

There is another allowed window for $\sigma$ (and $m_\phi$), in the pole region, where $m_\phi \simeq \frac{m_h}{2}$.
Dark matter with a mass half that of the Higgs has its annihilation process amplified by the exchange of an almost onshell Higgs boson.
The annihilation cross-section, summing over all Standard Model final states can be written as
\cite{Burgess:2000yq,Cline:2013gha}
\beq
\langle \sigma v \rangle 
= ~ \frac{8 \sigma^2 v^2}{(s-m_h^2)^2+m_h^2\Gamma_h^2} \cdot \frac{\Gamma_h}{m_h}
\simeq  \frac{8\sigma^2 v^2}{\Gamma_h m_h^3} 
\simeq 6.1 \times 10^{-9}  \left(\frac{\sigma}{10^{-5}} \right)^2~\rm{GeV^{-2}} \, ,
\label{Eq:sigmav}
\eeq
where $v\simeq 246$ GeV is the vacuum expectation value of the Higgs boson
and $\Gamma_h=4.07$ MeV its total width in the Standard Model.

Inserting (\ref{Eq:sigmav}) in Eq.(\ref{Eq:omega}), we obtain
\beq
\Omega_\phi h^2 \simeq 1.6 \times 10^8 \left(\frac{g_0}{g_{FO}} \right)
\sqrt{\frac{\alpha}{3}}\frac{15~\Gamma_h~m_h^3}{4 \sigma^2 v^2 M_P }
\simeq 0.05 ~\left( \frac{10^{-5}}{\sigma}\right    )^2,
\label{Eq:omegabis}
\eeq
where the masses are expressed in GeV and we took $g_{FO}=\frac{303}{4}$. 
We note, however, that the simple integration used to obtain Eq.~(\ref{Eq:omega}), fails badly in the region near the pole, particularly when 
$\Gamma_h/m_h$ is very small (as it is) \cite{Griest:1990kh}. Indeed, on the pole, the relic density is saturated when $\sigma \simeq 10^{-4}$ (see e.g. \cite{Cline:2013gha,Duerr:2015mva}). 
At $m_\phi = 62.6$ GeV, the XENON1T limit gives
$\sigma_{\phi p} \lesssim 6.5\times 10^{-47}~\rm{cm^2}$ which implies from
Eq.(\ref{Eq:sigmaphip}) $\sigma \lesssim 4 \times 10^{-3}$. This region $10^{-4}\lesssim \sigma \lesssim 4\times 10^{-3}$ corresponds to the purple region in Fig. \ref{fig:trhvssigma}. 

For an inflaton mass lighter than $\frac{m_h}{2}$, one must also take into
account the constraints on the invisible width of the Higgs 
given by the LHC. Indeed, decay processes $h \rightarrow \phi \phi$
is then allowed, with an invisible width
\beq
\Gamma(h \rightarrow \phi \phi) = \frac{\sigma^2 v^2}{4 \pi m_h} \sqrt{1-\frac{4 m_\phi^2}{m_h^2}} \, .
\label{Eq:gammah}
\eeq
The LHC Run2 limit from a combined ATLAS + CMS analysis gives \cite{Cepeda:2019klc}
\beq
Br(h \rightarrow {\rm invisible}) \lesssim 11\%.
\eeq
One can then extract from (\ref{Eq:gammah}) a limit on $\sigma$ for $m_\phi \lesssim \frac{m_h}{2}$ :
\beq
\sigma \lesssim 3.6\times 10^{-3},
\label{invlim}
\eeq
values for which the relic abundance dominated by the annihilation
to $b \bar b$ final state with
\beq
\langle \sigma v\rangle \sim 8 \sigma^2 \frac{v^2~\Gamma_h}{m_h^5} 
~~\Rightarrow ~~\frac{\Omega_\phi h^2}{0.1}  \sim \left(\frac{0.22}{\sigma}\right)^2 \, ,
\eeq
and indicates that if one satisfies the constraint from the invisible width of the Higgs (\ref{invlim}), there is a large over-density of dark matter when $m_\phi < m_h/2$ 
and away from the pole. 

One of the general consequences of the WIMPflation scenario is a relatively large coupling between the inflaton and Higgs ($\sigma \gtrsim 10^{-4}$) leading to a relatively large reheat temperature as we saw in the previous section.  If this model is derived from a supersymmetric/supergravity framework, gravitino production can be problematic \cite{prob,sw,eln,ego}. Gravitinos are produced during reheating and the gravitino abundance is typically proportional to the reheat temperature \cite{nos,ehnos,kl,ekn,Kawasaki:1994af,Moroi:1995fs,enor,Giudice:1999am,bbb,cefo,ps2,rs,egnop,Garcia:2017tuj,Garcia:2018wtq,Eberl:2020fml}. To ensure that gravitinos are not over-produced, one can derive a limit $T_{\rm RH} \lesssim 10^{10}$ GeV, far below the expected reheat temperature for large $\sigma$.

There are several possible solutions to the excessive gravitino abundance produced by a high reheating temperature. If the gravitino decays before the thermal freeze-out of the inflaton, then its energy density is reabsorbed in the thermal bath. This, however, requires a gravitino mass in excess of $\mathcal{O}(10^8)$ GeV, assuming a decay rate $\Gamma_{3/2} \propto m_{3/2}^3/M_P^2$. 
Alternatively, a modest amount of entropy production ($\mathcal{O}(10^4-10^5)$) injected before freeze-out would dilute the gravitino abundance to an acceptable level without over-diluting the baryon asymmetry. In both cases,
since we presume here that the inflaton is the dark matter, either the lightest supersymmetric particle must be under-abundant, or itself decay through $R$-parity violating interactions. 

\section{Conclusions}
\label{sec:concl}

There is an economic attractiveness in associating the inflaton with dark matter.
A simple potential for the inflaton containing a mass term, a quartic self-interaction, and a coupling to the Standard Model is sufficient for both inflation and dark matter. Phenomenologically viable models of inflation driven by quartic term
are easily constructed from $\alpha$-attractor T-models \cite{Kallosh:2013hoa} or Starobinsky-like models in the context of no-scale supergravity \cite{Garcia:2020eof,building}. In both examples, the quartic coupling 
is determined by the amplitude of large scale density fluctuations. Furthermore,
as we have argued, it is quite natural that the inflaton is stable in the no-scale
constructions \cite{ekoty,EGNO4}. 

Reheating and creating a thermal bath of radiation is an essential component of any inflationary model. Reheating is most commonly achieved through inflaton decays
to Standard Model particles. However, when inflaton oscillations are driven 
by a quartic (or higher-order) term, rather than a quadratic term, scatterings such as
$\phi \phi \to h h$ can also create a thermal bath, particularly if the $\phi^2 h^2$ coupling, $\sigma$ is large. We emphasize that inflation models driven by a quadratic term
(as in the Starobinsky model) can not reheat solely through scatterings.
While a thermal bath is created, it does not dominate the Universe at late times,
if the inflaton is stable. Similarly, a period of preheating does not lead to late-time
radiation domination for a stable inflaton.  In contrast, oscillations driven by quartic,
or higher-order terms, do lead to radiation domination. Contrary to na\"ive expectations,
reheating occurs for large $\sigma$, despite the large effective mass generated for the final state bosons.  While scatterings are suppressed due to kinematic effects, the suppression is power-law rather than exponential. 

We have also shown that 
for large values of $\sigma \gtrsim 10^{-3}$, reheating is almost instantaneous, and occurs during the first oscillation of the inflaton.  As a result, nearly all of the energy density stored in the inflaton potential is converted to radiation, leading to a relatively high reheat temperature in excess of $10^{14}$ GeV.   This may cause an
over-density of gravitinos, if the gravitinos are long-lived and there is no additional source of entropy production. 

A large value of $\sigma$ brings the inflaton back into equilibrium with the 
thermal bath. Absent of decays, the inflaton number density must be reduced
through annihilations as in WIMP thermal freeze-out scenarios. Indeed, the inflaton
at this stage acts as a scalar singlet dark matter candidate with Higgs portal 
couplings. It is well known that there remain two surviving mass ranges for 
inflaton dark matter. From the relic density of cold dark matter \cite{Planck} and the 
limits from XENON1T \cite{XENON}, the inflaton mass must be larger than  $\simeq 1.4$ TeV. Perturbativity of the couplings further limits the mass range to $\lesssim 5$ TeV.
In this case, $\sigma$ is of order a few tenths. 
Alternatively, the inflaton mass may be $\simeq m_h/2$ and annihilations are enhanced through the Higgs resonance. In this case, from direct detection limits, $\sigma$ must somewhat smaller $\lesssim 4 \times 10^{-3}$.  Scalar singlet dark matter and hence
WIMPflation can be tested in the next round of direct detection experiments.

\section*{Acknowledgments}
This project has received support from the European Union’s Horizon 2020 research and innovation programme under the Marie Sk$\lslash$odowska-Curie grant agreement No 860881-HIDDeN, the CNRS PICS MicroDark and the IN2P3 master projet "UCMN".
The work of K.A.O.~was supported in part by DOE grant DE-SC0011842  at the University of
Minnesota. The work of M.G. was supported by the Spanish Agencia Estatal de Investigaci\'on through Grants No.~FPA2015-65929-P (MINECO/FEDER, UE) and No.~PGC2018095161-B-I00, IFT Centro de Excelencia Severo Ochoa SEV-2016-0597, and Red Consolider MultiDark FPA2017-90566-REDC.

\end{document}